\documentclass[aps,twocolumn,showpacs,preprintnumbers,amsmath,amssymb,nofootinbib,superscriptaddress,showkeys]{revtex4}
\usepackage{hyperref}
\usepackage{float}
\usepackage{color}
\usepackage{epsfig}
\usepackage{graphicx}
\usepackage{mathptmx}      
\usepackage{latexsym}
\usepackage{longtable}
\usepackage{dcolumn}
\usepackage{natbib}
\usepackage[utf8]{inputenc}

\begin{document}

\title{Statistical error propagation in \emph{ab initio} no-core full
  configuration calculations of light nuclei}

\author{R. Navarro P\'erez}\email{navarroperez1@llnl.gov}
\affiliation{Nuclear and Chemical Science Division, Lawrence Livermore
  National Laboratory, Livermore, CA 94551, USA}

\author{J.E. Amaro}\email{amaro@ugr.es} \affiliation{Departamento de
  F\'{\i}sica At\'omica, Molecular y Nuclear \\ and Instituto Carlos I
  de F{\'\i}sica Te\'orica y Computacional \\ Universidad de Granada,
  E-18071 Granada, Spain.}
  
\author{E. Ruiz
  Arriola}\email{earriola@ugr.es} \affiliation{Departamento de
  F\'{\i}sica At\'omica, Molecular y Nuclear \\ and Instituto Carlos I
  de F{\'\i}sica Te\'orica y Computacional \\ Universidad de Granada,
  E-18071 Granada, Spain.} 

\author{P. Maris}\email{pmaris@iastate.edu} \affiliation{Department of
  Physics and Astronomy, Iowa State University, Ames, Iowa 50011, USA}

\author{J. P. Vary}\email{jvary@iastate.edu} \affiliation{Department
  of Physics and Astronomy, Iowa State University, Ames, Iowa 50011,
  USA}

\date{\today}

\begin{abstract} 
\rule{0ex}{3ex} We propagate the statistical uncertainty of
experimental NN scattering data into the binding energy of $^3$H and
$^4$He. We also study the sensitivity of the magnetic moment and
proton radius of the $^3$H to changes in the NN interaction. The
calculations are made with the no-core full configuration method in a
sufficiently large harmonic oscillator basis. For those light nuclei
we obtain $\Delta E^{\rm stat} (^3$H$)=0.015$ MeV and $\Delta E^{\rm
  stat} (^4$He$)=0.055$ MeV.
\end{abstract}

\maketitle

\section{Introduction}

The quantification of uncertainties in nuclear physics has gained a
significant interest in recent years~\cite{Dobaczewski:2014jga,
  McDonnell:2015sja, Carlsson:2015vda, Perez:2014bua}. In fact, a
special issue of the Journal of Physics G was dedicated entirely to
this particular topic~\cite{Ireland:0954}. Of course, the concentrated
attention is fully justified and the importance of providing
theoretical estimates with error bars cannot be overemphasized for
several reasons.  First, it provides guidance on the relation between
the input and the output along with the quantification of the
agreement/disagreement between theory and experiment. Second, it sets
the standards for the predictive power of the theory. Finally, it
helps to determine a balanced experimental program
by providing feedback from theory on what data are significant for
determining critical aspects of the theory.

The very idea of predictive power is unavoidably related to specifying
the links between the input and the output of the calculation.  In
this paper we are concerned with the impact of the uncertainties of
Nucleon-Nucleon (NN) interactions on the structure of the lightest
nuclei with $A=2,3,4$.  The implementation of such a program requires
a scrupulous determination of errors in NN potentials, for which the
main source of information has traditionally been the abundant
scattering experiments carried up to about the pion production
threshold over the last 70 years. In the case of NN scattering,
measurements are performed by counting events, giving a Poisson's
distribution to the measured observables. However, if the number of
events is large enough a standard normal distribution can be safely
assumed and experimental error bars can be given as a symmetric
$1\sigma$ confidence interval. Phenomenological NN potentials assume a
specific form for the interaction and adjust a number of parameters to
describe a collection of experimental scattering data. The parameter
fitting is usually done via a least squares procedure. Since each
experimental datum is provided with an error bar, the parameter space
contains a confidence region, rather than a single point, where the
description of data is considered correct. This region constitutes the
\emph{statistical} uncertainty for a particular NN
interaction. Different forms are assumed for the phenomenological NN
potentials describing correctly the same NN scattering database. These
differences are a source of \emph{systematic} uncertainty. A clear
sign of this systematic uncertainty is the incompatible predictions in
scattering observables for kinematic regions that have not yet been
measured~\cite{NavarroPerez:2012qf}. Another clear sign of this
systematic uncertainty is the different binding energies and other
observables predicted for light nuclei.  These differences in
properties of finite nuclei are attributable to many-body forces
which, in principle, are different for each NN interaction.

We address the specific situation where an NN interaction alone is
adopted as an input to make predictions on various observables of
light and heavy nuclei. In most cases the NN interaction is taken as
exact and only numerical or implementation errors are considered when
providing such calculations with an uncertainty. However, both
statistical and systematic errors are inherent to phenomenological
interactions and should be taken into account when quantifying the
total uncertainty of a nuclear structure calculation. For light nuclei
the no-core full configuration (NCFC) method~\cite{Maris:2008ax}
provides an \emph{ab initio} approach for extrapolating to an infinite
basis expansion of the nuclear wavefunction from a small set of finite
basis expansions with increasing size. The purpose of this work is to
use the NCFC approach to generate a realistic estimate of the
statistical uncertainty stemming from the experimental NN scattering
data, leaving the more complex propagation of systematic uncertainties
for a future endeavor.

The paper is organized as follows: section \ref{sec:NNPotential}
describes the major characteristics of the sum of Gaussians potential
which is the phenomenological NN interaction used in these
calculations. The theoretical framework to propagate the statistical
uncertainties is explained in section \ref{sec:Statistical}. The NCFC
method and the corresponding extrapolation uncertainty are shown in
section \ref{sec:NCFC}. Deuteron properties are calculated in section
\ref{sec:Deuteron} in order to test the convergence of the No-Core
Shell Model (NCSM)~\cite{Barrett:2013nh} with the sum of Gaussians
potential. Section \ref{sec:TritonAlfa} shows the NCFC calculations of
$^3$H and $^4$He binding energy with extrapolation and statistical
uncertainties. Finally, our conclusions are summarized in section
\ref{sec:Conlcusions}.

\section{Description of the NN potential}
\label{sec:NNPotential}

The NN interaction used for these nuclear structure calculations is
written as the sum of Gaussians potential and was introduced
in~\cite{Perez:2014yla}. This phenomenological interaction has been
fitted to the self-consistent Granada database with a total of 2995
neutron-proton and 3717 proton-proton experimental scattering
data~\cite{Perez:2013jpa}. The residuals of the potential (see section
\ref{sec:Statistical}) have been shown to follow the standard normal
distribution. This allows one to confidently propagate the experimental
uncertainty according to this distribution.

The potential has a clear boundary at $r_c = 3.0$ fm separating the short
and long range part of the NN interaction by means of
\begin{equation}
V(r) = V_{\rm short}(r)\theta(r_c - r) + V_{\rm long}(r)\theta(r-r_c).
 \label{eq:FullSOG_Potential}
\end{equation}
This differentiates between the purely phenomenological short range
part and the field theoretical long range tail which ensures the
proper analytic behavior of the scattering amplitude. The short range
part consists of a sum of Gaussian functions centered around the
origin in an operator basis
\begin{equation}
 V_{\rm short} (r) = \sum_{n=1}^{21} \hat{O}_n \left[ \sum_{i=1}^{4}
   V_{i,n} e^{-\frac{1}{2}\left( \frac{r(1+i)}{a} \right)^2} \right].
 \label{eq:SOG_Potential}
\end{equation}
The operators $\hat{O}_n$ correspond the AV18 basis plus three
additional operators introduced in Appendix A of~\cite{Perez:2013jpa},
the strength coefficients $V_{i,n}$ and width parameter $a$ are fitted
to describe the self consistent data base of~\cite{Perez:2013jpa}. The
values of the fitted parameters can be found in Table VI
of~\cite{Perez:2014yla}. The long range part includes the well known
charge dependent one-pion exchange (OPE) potential plus
electromagnetic interactions
\begin{equation}
  V_{\rm long}(r) = V_{\rm OPE}(r) + V_{\rm EM}(r).
\end{equation}
We refer to the full interaction in
  Eq(\ref{eq:FullSOG_Potential}) as the Gauss-OPE potential.

\section{Propagation of statistical uncertainties}
\label{sec:Statistical}

A chi square minimization scheme corresponds to the trial assumption
that the fitted data are independent and normally distributed. When
the fitting model is flexible enough to accurately describe the data,
this assumption becomes equivalent to the assumption that the
discrepancies between theory and experiment follow the standard
normal distribution i.e.
\begin{equation}
 R_i = \frac{O_i^{\rm exp}-O_i^{\rm theor}}{\Delta O_i^{\rm exp}} \sim
 N(0,1),
\end{equation}
where $R_i$ are known as residuals. If the resulting residuals of a
fit do not follow the standard normal distribution the trial
assumption is invalid and any subsequent error propagation based on
normality is at the very least questionable, if not wrong. For the
case of the sum of Gaussians potential the normality of residuals has
been stringently tested and confirmed~\cite{Perez:2014yla}. The
fitting procedure used to adjust the potential parameters allows one
to estimate the corresponding covariance matrix $\cal C$, which
propagates the experimental error bars of the fitted data to the
parameters and assumes a multivariate normal distribution for
them. The covariance matrix is defined as the inverse of the Hessian
matrix $H$
\begin{equation}
 ({\cal C}^{-1})_{i j} \equiv H_{i j} = \frac{\partial
    \chi^2}{\partial p_i \partial p_j},
\end{equation}
where $\chi^2 = \sum_i R_i^2$ and $\mathbf{p}$ is the vector
containing the fitting parameters. To propagate the statistical
uncertainties from the interaction into nuclear structure calculations
one could, in principle, directly use the covariance matrix with
\begin{equation}
 (\Delta F)^2 = \sum_{i j} \frac{\partial F}{\partial
    p_i}\frac{\partial F}{\partial p_j} {\cal C}_{i j,}
  \label{eq:CovarianceError}
\end{equation}
where $F$ is any quantity calculated as a function of the potential
parameters. However, the level of complexity in most of the nuclear
structure numerical methods makes the calculation of the derivatives
in Eq.~(\ref{eq:CovarianceError}) computationally costly; a simple
finite differences method at fourth order would require one to make
four evaluations of $F$ to calculate the derivative with respect to
each parameter. Since the sum of Gaussians potential has 44
independent fitting parameters~\footnote{Although
  Eq.(\ref{eq:SOG_Potential}) indicates a total of $21 \times 4 + 1 =
  85$ fitting parameters, some of them correspond to linear
  combinations of others while some others are fixed to zero.}, a total of
176 evaluations would be necessary to propagate the statistical
uncertainty to a single quantity. While one could alternatively
proceed to differentiate the NN interaction analytically, term by
term, and evaluate the single terms separately, we will proceed
differently here.

Monte-Carlo techniques provide an efficient approach to study the
sensitivity of elaborate nuclear structure calculations to variations
on the input interaction. To generate a Monte-Carlo family of
potential parameters one simply has to draw random numbers following a
multivariate normal distribution according to the covariance matrix
$\cal C$
\begin{equation}
 P(p_1,p_2,\ldots,p_P) = \frac{1}{\sqrt{(2\pi)^P {\rm det} {\cal C}}}
 e^{-\frac{1}{2}(\mathbf p -\mathbf p_0)^T{\cal C} (\mathbf p
   -\mathbf p_0)},
\end{equation}
where $\mathbf p_0$ has the central values of the fitted
parameters. This sampling can be easily done by taking advantage of
the Cholesky decomposition ${\cal C = L L}^T$ and using independent
standard normal variates $\mathbf z = (z_1, z_2, \ldots, z_p)$ with
\begin{equation}
 \mathbf p = \mathbf p_0 + {\cal L} \cdot \mathbf z.
\end{equation}
With $M$ different samples of $\mathbf p$ one can directly make $M$
evaluations of $F$ and see the resulting spread and distribution.
This technique has been recently employed in~\cite{Perez:2014laa} to
estimate the statistical uncertainty of the Triton ground state
energy. That particular study was done with $M=205$, however a much
smaller sample of $M=30$ already gives a fairly similar estimate. A
direct bootstrapping of the experimental data has also been used to
propagate statistical uncertainties stemming from NN scattering
data~\cite{Perez:2014jsa}. The use of a reduced sample has also been
checked using larger populations both for triton and the alpha
particle using the Faddeev-Yakubovsky equations~\cite{Nogga-2015}.

\section{No-core full configuration method}
\label{sec:NCFC}

The No-Core Full Configuration (NCFC) method, as an \emph{ab initio}
approach, solves the many body Schr\"odinger equation 
\begin{eqnarray}
  H \; \Psi_i(\vec{r}_1, \ldots, \vec{r}_A) &=& 
     E_i \; \Psi_i(\vec{r}_1, \ldots, \vec{r}_A)
\label{eq:eigenvalueproblem}
\end{eqnarray}
by expanding the corresponding wavefunction of $Z$ protons and $N$
neutrons in a $A=Z+N$-body basis of Slater determinants $\Phi_k$ of
single-particle wave functions $\phi_{nljm}(\vec{r})$
\begin{eqnarray}
  \Psi(\vec{r}_1, \ldots, \vec{r}_A) = 
    \sum c_k \Phi_k(\vec{r}_1, \ldots, \vec{r}_A) \,,
\label{eq:expansion}
\end{eqnarray}
with 
$$ \Phi_k(\vec{r}_1, \ldots, \vec{r}_A) = {\cal A} [ \phi_{n_1 l_1 j_1
    m_1}(\vec{r}_1) \, \phi_{n_2 l_2 j_2 m_2}(\vec{r}_2) \ldots
  \phi_{n_A l_A j_A m_A}(\vec{r}_A) ]
$$ 
and ${\cal A}$ the antisymmetrization operation. Even though it is
customary to use the harmonic-oscillator (HO) basis for the
single-particle wavefunctions, the method can be extended to more
general single-particle bases~\cite{Caprio:2012rv}. The
single-particle wavefunction labels indicate the quantum numbers: $n$
and $l$ are the radial and orbital HO quantum numbers (with $N_i =
2n_i + l_i$ the number of HO quanta), $j$ is the total spin and $m$
is its projection along the $z$-axis. Once the basis expansion of the
many-body wavefunction has been made Eq.~(\ref{eq:eigenvalueproblem})
becomes a linear algebra eigenvalue problem with sparse matrices. The
many-body Hamiltonian $H$ in Eq.~(\ref{eq:eigenvalueproblem}) can be
expressed in terms of the relative kinetic energy plus 2-body, 3-body,
and, in general, up to $A$-body interaction terms
\begin{eqnarray}
  H &=& T_{\hbox{\scriptsize rel}} 
    + V_{\hbox{\scriptsize Coulomb}}
    + V_{\hbox{\scriptsize NN}} + V_{\hbox{\scriptsize NNN}} + \ldots
\label{eq:hamiltonian}
\end{eqnarray}
In this work we restrict ourselves to a 2-body (NN) interaction,
leaving the inclusion of appropriate 3 and more body forces to future
investigations. 

Of course the numerical solution of the eigenvalue problem in
Eq.~(\ref{eq:eigenvalueproblem}) requires the truncation of the
infinite-dimensional basis expansion. Because of this truncation, the
solution gives a strict upper bound for the lowest state with a given
spin and parity. The NCFC method establishes the convergence pattern
as a functions of the HO energy along with increasing basis space
dimension and extrapolates to a complete infinite basis. This requires
the solution of eigenvalue problems of considerably large matrices
with dimensions well over a billion. Therefore the algorithms used to
construct and operate with these matrices are required to make an
efficient use of the available computational
resources~\cite{sternberg2008accelerating, maris2010scaling,
  aktulga2012topology}

We use the $N_{\rm max}$ truncation, which restricts the total number
of HO quanta of the many-body basis: the basis is limited to many-body
states with $\sum_{A} N_i \le N_0 + N_{\rm max}$, where $N_i$ is the
number of quanta of each single-particle state; $N_0$ is the minimal
number of quanta for that nucleus; and $N_{\rm max}$ is the truncation
parameter.  For HO single-particle states, this truncation leads to an
exact factorization of the center-of-mass wave function and the
relative wave function~\cite{Navratil:2009ut, Barrett:2013nh,
  Caprio:2012rv, gloeckner1974spurious}

%
\subsection{Extrapolation Method and Extrapolation Uncertainty Quantification}
\label{sec:extrunc} 
We use the empirical extrapolation introduced in~\cite{Forssen:2008qp}
and expanded in~\cite{Maris:2013poa,Maris:2008ax} for the ground state
energy at a fixed value of the oscillator constant $\hbar \Omega$
\begin{equation}
 E_{\rm gs} (N_{\rm max})=a\exp(-c N_{\rm max})+E_{\infty}.
 \label{eq:Expfit}
\end{equation}
This type of extrapolation requires one to solve
Eq.(~\ref{eq:eigenvalueproblem}) multiple times to calculate $E_{\rm
  gs}$ with at least three different values of the cut-off parameter 
$N_{\rm max}$ in order to fit the $a$, $c$ and $E_\infty$  
parameters. An estimate of the solution of 
Eq.(~\ref{eq:eigenvalueproblem}) with a infinitely large basis is 
given by $E_\infty$. 

Given the variational nature of this approach an extrapolation can be
improved, by either fitting the parameters in Eq.(~\ref{eq:Expfit}) to
a larger number of points or by fitting to ground state energies with
larger basis size. With this in mind the extrapolation uncertainty is
quantified by direct comparison with extrapolations based on lower
$N_{\rm max}$ results.  In~\cite{Maris:2008ax,Maris:2013poa} an
extrapolation is made by taking sets of three $E(N_{\rm max})$ values,
say $N_{\rm max}=10, \ 12, \ 14$ while the extrapolation uncertainty
is estimated by taking the difference with the extrapolation using
$N_{\rm max} = 8, \ 10, \ 12$. For this work we perform
all extrapolations by starting with $N_{\rm max}=8$
  and include consecutive even values up to a certain number of points
  $N_p$. The extrapolation uncertainty is estimated by comparing with
an extrapolation using one point less. For example, an extrapolation
with $N_p=4$ uses $N_{\rm max} = 8, \ 10, \ 12,
\ 14$ and the corresponding uncertainty is the difference with the
extrapolation using the three points $N_{\rm max} = 8, 10, 12 $. These
extrapolations with a larger number of points can reduce the
extrapolation uncertainty up to two orders of magnitude while keeping
the statistical uncertainty within the same order. Also, this approach
is self-consistent in the sense that the extrapolation uncertainty
decreases as the number of points increases and for a given number of
points the extrapolation is within the uncertainty of smaller
extrapolations. Although additional methods are available for
extrapolating to the ground state energy with an infinite basis from
calculations using truncated HO bases~\cite{Coon:2012ab,
  Furnstahl:2012qg}, the method described here allows us to compare
the size of the extrapolation error with the statistical uncertainty.

\section{Deuteron and convergence rate}
\label{sec:Deuteron}

\begin{figure}
\centering
\vskip2ex
\includegraphics[width=\linewidth]{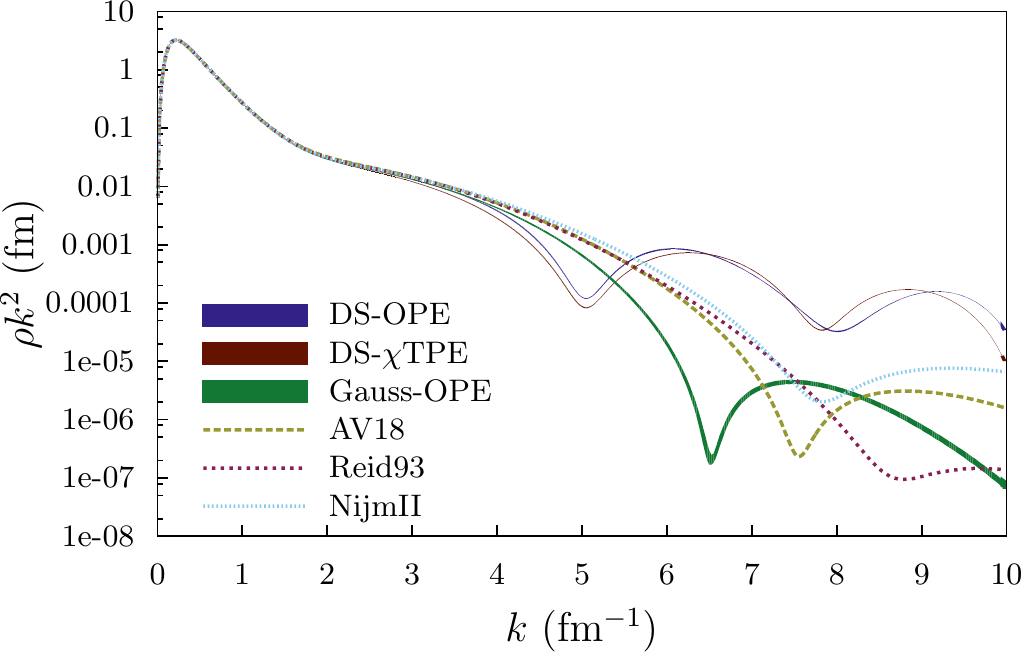}
\caption{(Color online) Deuteron momentum distribution for different
  local potentials. DS-OPE (blue band), DS-$\chi$TPE (red band),
  Gauss-OPE (green band), AV18 (yellow dashed line), Reid93 (red
  short-dashed line) and NijmII (light blue dotted line).}
\label{fig:2HMomemtumDist}       
\end{figure}

In order to get an initial insight into the convergence of NCSM
calculations using the Gauss-OPE potential we calculated first the
Deuteron momentum distribution using different phenomenological
potentials. In particular, we use ArgonneV18~\cite{Wiringa:1994wb},
Reid93~\cite{Stoks:1994wp},
NijmII~\cite{Stoks:1994wp}, and the delta shell potentials
DS-OPE~\cite{Perez:2013jpa,Perez:2013mwa} and
DS-$\chi$TPE~\cite{Perez:2013oba}, in addition to the Gauss-OPE
potential~\cite{Perez:2014yla}. For the DS-OPE, DS-$\chi$TPE and
Gauss-OPE potentials, we included the results from the Monte-Carlo
sampling described in the previous sections to form bands of
results. The momentum distributions can be seen in
Fig.~\ref{fig:2HMomemtumDist}. It is remarkable that the delta-shell
potentials contain the largest high-momentum component despite having
no repulsive hard core. This is probably due to the discontinuities in
which the potential goes from zero to infinity. In fact, these large
momentum distributions become manifest when performing NCSM
calculations with these two interactions as the convergence rate is
significantly slower. Out of the remaining interactions, the Gauss-OPE
and Reid93 potentials give the smallest high momentum contribution
indicating that they are considerably softer than the other potentials
considered. We also note in passing the clearly incompatible high
momentum distributions even within the 1$\sigma$ statistical
uncertainty shown as a band for the DS-OPE, DS-$\chi$TPE and
Gauss-OPE. These incompatibilities are indeed expected and are a
signal of the present systematic uncertainties which still need to be
quantified.

\begin{figure*}
\centering
\includegraphics[width=\columnwidth]{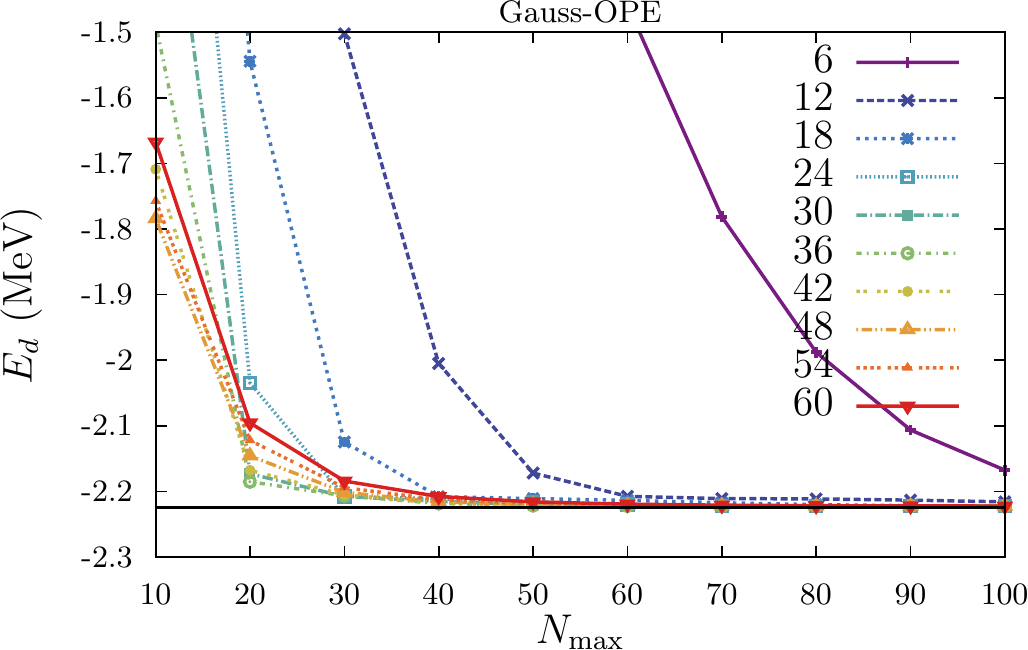}
\includegraphics[width=\columnwidth]{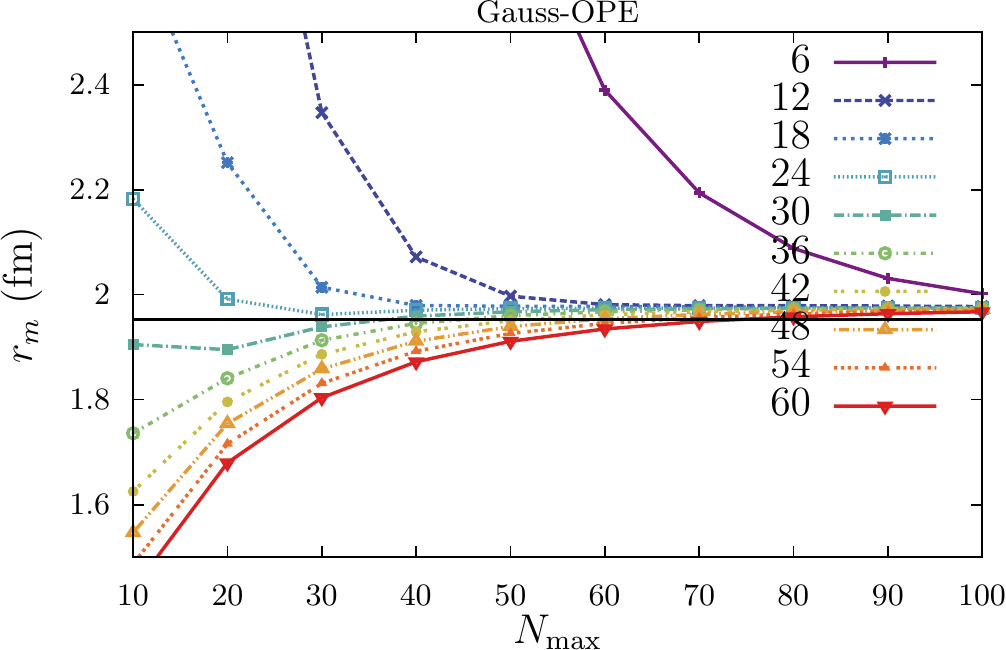}
\vskip2ex
\includegraphics[width=\columnwidth]{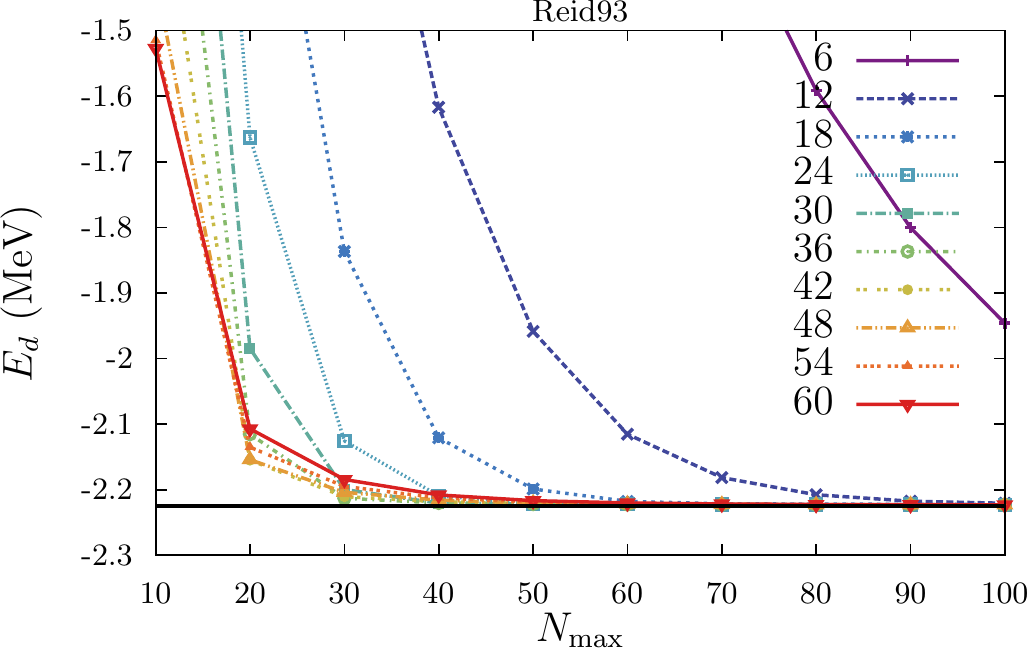}
\includegraphics[width=\columnwidth]{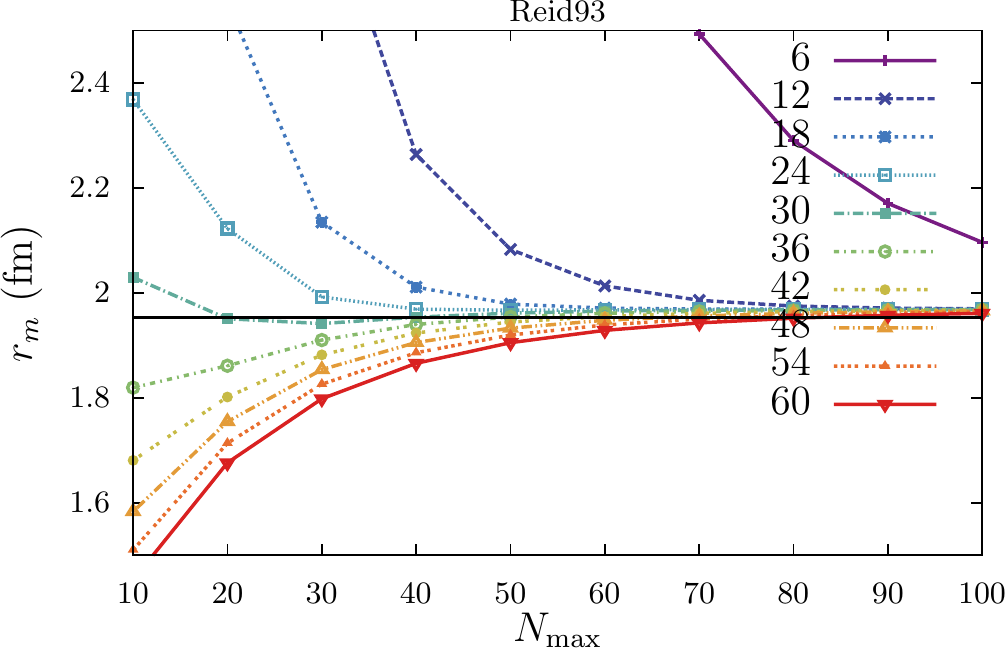}
\caption{(Color online) NCSM calculations of the Deuteron binding
  energy (left panels) and root mean square radius (right panels) as a
  function of the basis truncation parameter $N_{\rm max}$ with
  different values of the harmonic oscillator parameter $\hbar \Omega$
  for the Gauss-OPE (top panels) and Reid93 (bottom panels)
  potentials. The Gauss-OPE potential shows a faster convergence rate
  in both quantities as evident from the smaller spread with respect
  to $\hbar \Omega$ at fixed $N_{\rm max}$.}
\label{fig:DeuteronConvergence}       
\end{figure*}

In order to directly test the convergence of NCSM calculations using
the Gauss-OPE potential, we calculated the binding
energy and root mean square radius of the Deuteron using different
values of the oscillator parameter $\hbar \Omega$ and different basis
truncation $N_{\rm max}$ and compare with the same calculations using
the Reid93 potential. The results are shown in Figure
\ref{fig:DeuteronConvergence}. As can be seen
Gauss-OPE has a faster convergence rate which can be
traced to its significantly softer core.  All of this is in agreement
with our previous Weinberg eigenvalue analyses of these and other
local interactions~\cite{Perez:2014bua}.

\section{NCFC calculations of $^3$H and $^4$He}
\label{sec:TritonAlfa}

Once Gauss-OPE has been shown to have an advantageous convergence rate
for {\it ab initio} NCSM calculations of light nuclei we use it to
propagate the statistical uncertainty stemming from experimental NN
scattering data into the calculation of the binding energy of $^3$H
and $^4$He. As a first step we looked for the variational minimum of
the ground state energy of $^3$H and $^4$He as a function of the
oscillator parameter $\hbar \Omega$ with the sufficiently large basis
size $N_{\rm max} = 20$ and the central values of the Gauss-OPE
potential parameters. The minima were found at $\hbar \Omega = 60$MeV
for $^3$H and $\hbar \Omega = 65$MeV for $^4$He. Having found the
variational minimum, the Monte-Carlo method described in
section~\ref{sec:Statistical} was used to generate a family of 33
interactions following the multivariate Gauss distribution of the
potential parameters befitting the covariance matrix. With each set of
parameters the ground state binding energy of $^3$H was calculated
using different many-body basis space cutoffs $N_{\rm max} =
8,10,12,14,16,18,20$ with the value of $\hbar \Omega$ fixed at the
variational minimum mentioned above. These evaluations allowed us to
perform, for sets of $N_{\rm max}$ values with three to seven entries,
33 extrapolations following Eq.(\ref{eq:Expfit}) to estimate the
corresponding statistical
uncertainty. Table~\ref{tab:ExpfitConvergence} shows the resulting
extrapolations for $^3$H and $^4$He with different number $N_{\rm
  max}$ entries; a clear convergence pattern can be seen as larger
basis spaces are included in the extrapolation for the complete space
estimate and the corresponding uncertainties. Other observables can be
extracted from the resulting many-body wavefunction of the truncated
oscillator basis expansion. In particular, we investigated the
magnetic moment $\mu$ and proton radius $r_p$ of $^3$H and their
sensitivity to the Monte-Carlo sampling of the interaction parameters
distribution. For these quantities we find $\mu = 2.610(1) \ \mu_N$
$r_p = 1.4668(8)$fm.

\begin{table*}
 \caption{\label{tab:ExpfitConvergence} $^3$H and $^4$He ground state
   energy NCFC extrapolation for the Gauss-OPE potential using
   different number of $E(N_{\rm max})$ points $N_p$ according to
   Eq.(~\ref{eq:Expfit}). All extrapolations start with $N_{\rm max} =
   8$ and include consecutive even values up to the corresponding
   number of points $N_p$. $E_\infty$ is the mean of the 33
   Monte-Carlo calculations with that number of points, the
   extrapolation uncertainty $\Delta E_\infty^{\rm ext}$ is the
   difference between the extrapolation with $N_p$ points listed and
   the extrapolation with a single point less. The statistical
   uncertainty $\Delta E_\infty^{\rm stat}$ is given by the standard
   deviation of the Monte-Carlo extrapolations with the indicated
   number of entries $N_p$. For comparison we show experimental values
   in the last line~\cite{Audi:2002rp}. Energies are in units of MeV}
 \begin{ruledtabular}
 \begin{tabular*}{\linewidth}{@{\extracolsep{\fill}} l c c c c c c    }
       & \multicolumn{3}{c}{$^3$H}  
       & \multicolumn{3}{c}{$^4$He} \\  
\cline{2-4} \cline{5-7} 
 $N_p$ & \multicolumn{1}{c}{$E_\infty$} 
       & \multicolumn{1}{c}{$\Delta E_\infty^{\rm ext}$} 
       & \multicolumn{1}{c}{$\Delta E_\infty^{\rm stat}$}
       & \multicolumn{1}{c}{$E_\infty$} 
       & \multicolumn{1}{c}{$\Delta E_\infty^{\rm ext}$} 
       & \multicolumn{1}{c}{$\Delta E_\infty^{\rm stat}$}  \\
 \hline \noalign{\smallskip}
 3 &  -8.289 &   $-$  & 0.061 &  -29.224 &   $-$  & 0.226 \\ 
 4 &  -7.408 & 0.881  & 0.020 &  -25.980 & 3.244  & 0.085 \\ 
 5 &  -7.486 & 0.078  & 0.019 &  -25.417 & 0.563  & 0.066 \\ 
 6 &  -7.421 & 0.065  & 0.016 &  -25.102 & 0.315  & 0.058 \\ 
 7 &  -7.424 & 0.003  & 0.015 &  -24.934 & 0.168  & 0.055 \\ 
 \hline \noalign{\smallskip}
 $E^{\rm Exp}$ & -8.482 & & & -28.296 & & \\
 \end{tabular*}
 \end{ruledtabular}
\end{table*}

\begin{figure*}
\centering
\vskip2ex
\includegraphics[width=\columnwidth]{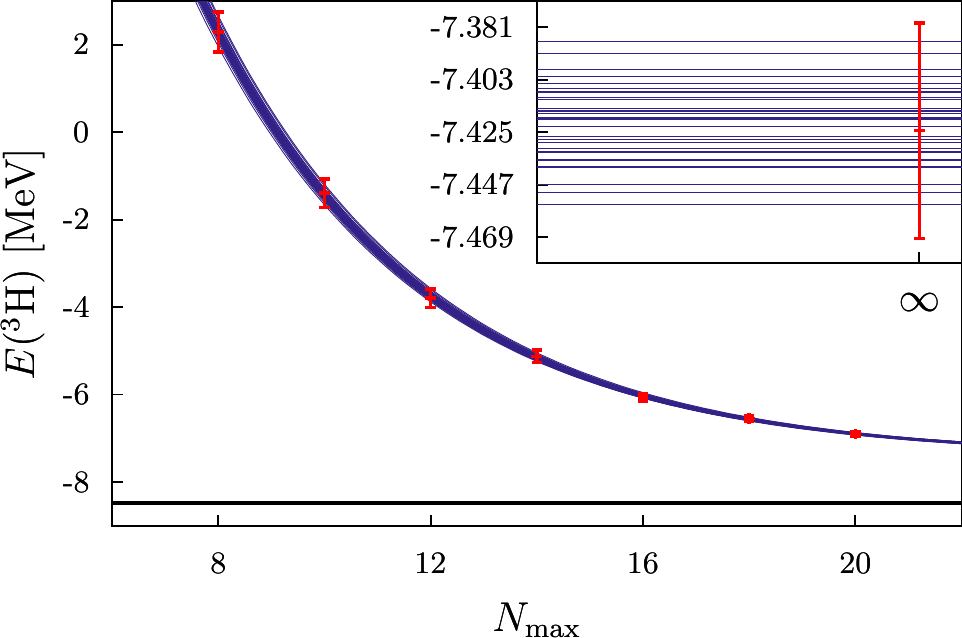} \ \
\includegraphics[width=\columnwidth]{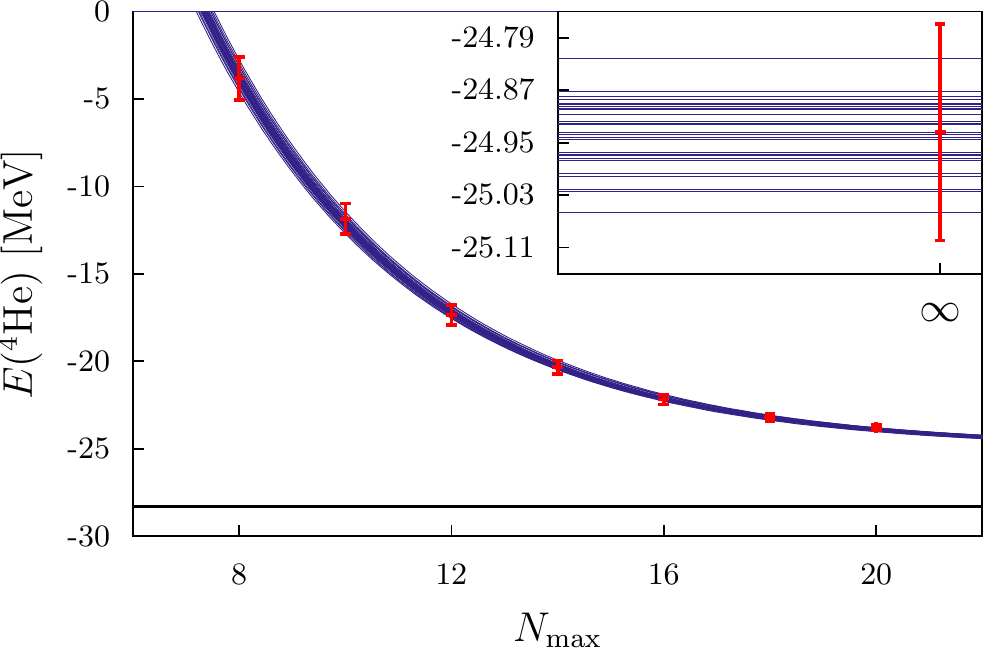}
\caption{(Color online) NCSM ground state energy as a function of
  $N_{\rm max}$ for $^3$H at $\hbar \Omega = 60$ MeV (left panel) and
  $^4$He at $\hbar \Omega = 65$ MeV (right panel) with the Gauss-OPE
  potential. The red error bars indicate the mean and 3$\sigma$
  confidence interval of the 33 calculations with different potential
  parameters (see main text). The blue solid curves give the
  corresponding 33 extrapolations fitting the $a$, $c$ and $E_\infty$
  of Eq.\ref{eq:Expfit} to the calculated values of $E(N_{\rm max} =
  8, 10, \ldots, 20)$. The panels at the top right corners show the
  extrapolated value at $N_{\rm max} \rightarrow \infty$ from each fit
  with an error bar indicating the corresponding mean and 3$\sigma$
  confidence interval. The solid black indicates the experimental
  value.}
\label{fig:3HExpFit}       
\end{figure*}

The extrapolations using the seven available $E(N_{\rm max})$ values
and those corresponding to the last row of
Table~\ref{tab:ExpfitConvergence} are shown in
Figure~\ref{fig:3HExpFit} as blue lines. As can be seen from the
finite basis calculations and the corresponding extrapolations the
statistical uncertainty decreases as the cutoff parameter $N_{\rm
  max}$ is increased.

We now call attention to the inserts in Figure~\ref{fig:3HExpFit}
where we display, on a greatly expanded scale, the extrapolated ground
state energies for each of the 33 Monte-Carlo samples.  The visual
impression is that these results are distributed in a manner that may
be consistent with a Gaussian distribution though we do not carry out
a detailed study of these limited distributions. These results have
some additional impact on the numerical method itself. While in the
case of the triton the dominating uncertainty for
$N_p=7$, using calculations up to $N_{\rm max} =
  20$, is the statistical one, in the alpha particle case the
situation is just the opposite, indicating that higher
$N_{\rm max}$ values should probably be pursued to
confirm the statistical uncertainty. Of course, in both situations
there is a mismatch with the experimental binding energies, for which
the traditional explanation rests on the inclusion of three- and
four-body forces.

In addition, we have restricted the analysis to the statistical
uncertainties of the NN interaction stemming directly from the
scattering data. In a previous work~\cite{Perez:2014waa} the role
played by the potential representation has been analyzed and it has
been found that there are statistically equivalent interactions which,
however, produce larger systematic uncertainties by about an order of
magnitude in scattering observables. Thus, we expect that when this
additional systematic uncertainty is included in the analysis the
current numerical uncertainty for $N_p=7$ will actually be small by
comparison.

\section{Conclusions}
\label{sec:Conlcusions}

We performed NCFC calculations of $^3$H and $^4$He combined with
Monte-Carlo techniques to extract realistic estimates of the
statistical uncertainty originating from uncertainties in experimental
NN scattering data. The calculations used the Gauss-OPE potential
which showed an appropriate convergence rate for calculations in light
nuclei given its soft core nature. The converged result from
extrapolation in $^3$H is in agreement with the previous findings
in~\cite{Perez:2014laa}. In all cases the uncertainty of the NCFC
extrapolation was assessed by comparing with extrapolations based on a
reduced number of entries for determining the fit function. For the
case of $^3$H it was possible to obtain a result where the
extrapolation uncertainty is smaller than the statistical
uncertainty. However, this was not possible for the $^4$He
case. Nonetheless one should keep in mind that the \emph{systematic}
uncertainty, according to preliminary studies~\cite{Perez:2014waa}, is
expected to be an order of magnitude larger than the statistical
uncertainty. Therefore, it may not be profitable to reduce the
extrapolation uncertainty by performing calculations with an even
larger basis size.

This work was supported by the US Department of Energy under Grants
No. DESC0008485 (SciDAC/NUCLEI) and No.  DE-FG02-87ER40371, by the US
National Science Foundation under Grant No. 0904782. Computational
resources were provided by the National Energy Research Supercomputer
Center (NERSC), which is supported by the Office of Science of the
U.S. Department of Energy under Contract No. DE-AC02-05CH11231. This
work was also supported by Spanish DGI (grant FIS2014-29386-P) and Junta
de Andaluc{\'{\i}a} (grant FQM225). This work was partly performed
under the auspices of the U.S. Department of Energy by Lawrence
Livermore National Laboratory under Contract No. DE-AC52-07NA27344.

\bibliography{biblio}

\begin{thebibliography}{29}
\expandafter\ifx\csname natexlab\endcsname\relax\def\natexlab#1{#1}\fi
\expandafter\ifx\csname bibnamefont\endcsname\relax
  \def\bibnamefont#1{#1}\fi
\expandafter\ifx\csname bibfnamefont\endcsname\relax
  \def\bibfnamefont#1{#1}\fi
\expandafter\ifx\csname citenamefont\endcsname\relax
  \def\citenamefont#1{#1}\fi
\expandafter\ifx\csname url\endcsname\relax
  \def\url#1{\texttt{#1}}\fi
\expandafter\ifx\csname urlprefix\endcsname\relax\def\urlprefix{URL }\fi
\providecommand{\bibinfo}[2]{#2}
\providecommand{\eprint}[2][]{\url{#2}}

\bibitem[{\citenamefont{Dobaczewski et~al.}(2014)\citenamefont{Dobaczewski,
  Nazarewicz, and Reinhard}}]{Dobaczewski:2014jga}
\bibinfo{author}{\bibfnamefont{J.}~\bibnamefont{Dobaczewski}},
  \bibinfo{author}{\bibfnamefont{W.}~\bibnamefont{Nazarewicz}},
  \bibnamefont{and} \bibinfo{author}{\bibfnamefont{P.~G.}
  \bibnamefont{Reinhard}}, \bibinfo{journal}{J. Phys.}
  \textbf{\bibinfo{volume}{G41}}, \bibinfo{pages}{074001}
  (\bibinfo{year}{2014}), \eprint{1402.4657}.

\bibitem[{\citenamefont{McDonnell et~al.}(2015)\citenamefont{McDonnell,
  Schunck, Higdon, Sarich, Wild, and Nazarewicz}}]{McDonnell:2015sja}
\bibinfo{author}{\bibfnamefont{J.~D.} \bibnamefont{McDonnell}},
  \bibinfo{author}{\bibfnamefont{N.}~\bibnamefont{Schunck}},
  \bibinfo{author}{\bibfnamefont{D.}~\bibnamefont{Higdon}},
  \bibinfo{author}{\bibfnamefont{J.}~\bibnamefont{Sarich}},
  \bibinfo{author}{\bibfnamefont{S.~M.} \bibnamefont{Wild}}, \bibnamefont{and}
  \bibinfo{author}{\bibfnamefont{W.}~\bibnamefont{Nazarewicz}},
  \bibinfo{journal}{Phys. Rev. Lett.} \textbf{\bibinfo{volume}{114}},
  \bibinfo{pages}{122501} (\bibinfo{year}{2015}), \eprint{1501.03572}.

\bibitem[{\citenamefont{Carlsson et~al.}(2015)\citenamefont{Carlsson, Ekström,
  Forssén, Strömberg, Lilja, Lindby, Mattsson, and Wendt}}]{Carlsson:2015vda}
\bibinfo{author}{\bibfnamefont{B.~D.} \bibnamefont{Carlsson}},
  \bibinfo{author}{\bibfnamefont{A.}~\bibnamefont{Ekström}},
  \bibinfo{author}{\bibfnamefont{C.}~\bibnamefont{Forssén}},
  \bibinfo{author}{\bibfnamefont{D.~F.} \bibnamefont{Strömberg}},
  \bibinfo{author}{\bibfnamefont{O.}~\bibnamefont{Lilja}},
  \bibinfo{author}{\bibfnamefont{M.}~\bibnamefont{Lindby}},
  \bibinfo{author}{\bibfnamefont{B.~A.} \bibnamefont{Mattsson}},
  \bibnamefont{and} \bibinfo{author}{\bibfnamefont{K.~A.} \bibnamefont{Wendt}}
  (\bibinfo{year}{2015}), \eprint{1506.02466}.

\bibitem[{\citenamefont{Navarro~Perez et~al.}(2015)\citenamefont{Navarro~Perez,
  Amaro, and Arriola}}]{Perez:2014bua}
\bibinfo{author}{\bibfnamefont{R.}~\bibnamefont{Navarro~Perez}},
  \bibinfo{author}{\bibfnamefont{J.~E.} \bibnamefont{Amaro}}, \bibnamefont{and}
  \bibinfo{author}{\bibfnamefont{E.~R.} \bibnamefont{Arriola}},
  \bibinfo{journal}{Phys. Rev.} \textbf{\bibinfo{volume}{C91}},
  \bibinfo{pages}{054002} (\bibinfo{year}{2015}), \eprint{1411.1212}.

\bibitem[{\citenamefont{Ireland and Nazarewicz}(2015)}]{Ireland:0954}
\bibinfo{author}{\bibfnamefont{D.~G.} \bibnamefont{Ireland}} \bibnamefont{and}
  \bibinfo{author}{\bibfnamefont{W.}~\bibnamefont{Nazarewicz}},
  \bibinfo{journal}{J. Phys.} \textbf{\bibinfo{volume}{G42}},
  \bibinfo{pages}{030301} (\bibinfo{year}{2015}).

\bibitem[{\citenamefont{Navarro~Pérez
  et~al.}(2013{\natexlab{a}})\citenamefont{Navarro~Pérez, Amaro, and
  Ruiz~Arriola}}]{NavarroPerez:2012qf}
\bibinfo{author}{\bibfnamefont{R.}~\bibnamefont{Navarro~Pérez}},
  \bibinfo{author}{\bibfnamefont{J.~E.} \bibnamefont{Amaro}}, \bibnamefont{and}
  \bibinfo{author}{\bibfnamefont{E.}~\bibnamefont{Ruiz~Arriola}},
  \bibinfo{journal}{Phys. Lett.} \textbf{\bibinfo{volume}{B724}},
  \bibinfo{pages}{138} (\bibinfo{year}{2013}{\natexlab{a}}),
  \eprint{1202.2689}.

\bibitem[{\citenamefont{Maris et~al.}(2009)\citenamefont{Maris, Vary, and
  Shirokov}}]{Maris:2008ax}
\bibinfo{author}{\bibfnamefont{P.}~\bibnamefont{Maris}},
  \bibinfo{author}{\bibfnamefont{J.}~\bibnamefont{Vary}}, \bibnamefont{and}
  \bibinfo{author}{\bibfnamefont{A.}~\bibnamefont{Shirokov}},
  \bibinfo{journal}{Phys.Rev.} \textbf{\bibinfo{volume}{C79}},
  \bibinfo{pages}{014308} (\bibinfo{year}{2009}), \eprint{0808.3420}.

\bibitem[{\citenamefont{Barrett et~al.}(2013)\citenamefont{Barrett, Navratil,
  and Vary}}]{Barrett:2013nh}
\bibinfo{author}{\bibfnamefont{B.~R.} \bibnamefont{Barrett}},
  \bibinfo{author}{\bibfnamefont{P.}~\bibnamefont{Navratil}}, \bibnamefont{and}
  \bibinfo{author}{\bibfnamefont{J.~P.} \bibnamefont{Vary}},
  \bibinfo{journal}{Prog.Part.Nucl.Phys.} \textbf{\bibinfo{volume}{69}},
  \bibinfo{pages}{131} (\bibinfo{year}{2013}).

\bibitem[{\citenamefont{Navarro~Perez
  et~al.}(2014{\natexlab{a}})\citenamefont{Navarro~Perez, Amaro, and
  Ruiz~Arriola}}]{Perez:2014yla}
\bibinfo{author}{\bibfnamefont{R.}~\bibnamefont{Navarro~Perez}},
  \bibinfo{author}{\bibfnamefont{J.}~\bibnamefont{Amaro}}, \bibnamefont{and}
  \bibinfo{author}{\bibfnamefont{E.}~\bibnamefont{Ruiz~Arriola}},
  \bibinfo{journal}{Phys.Rev.} \textbf{\bibinfo{volume}{C89}},
  \bibinfo{pages}{064006} (\bibinfo{year}{2014}{\natexlab{a}}),
  \eprint{1404.0314}.

\bibitem[{\citenamefont{Navarro~Pérez
  et~al.}(2013{\natexlab{b}})\citenamefont{Navarro~Pérez, Amaro, and
  Ruiz~Arriola}}]{Perez:2013jpa}
\bibinfo{author}{\bibfnamefont{R.}~\bibnamefont{Navarro~Pérez}},
  \bibinfo{author}{\bibfnamefont{J.}~\bibnamefont{Amaro}}, \bibnamefont{and}
  \bibinfo{author}{\bibfnamefont{E.}~\bibnamefont{Ruiz~Arriola}},
  \bibinfo{journal}{Phys.Rev.} \textbf{\bibinfo{volume}{C88}},
  \bibinfo{pages}{064002} (\bibinfo{year}{2013}{\natexlab{b}}),
  \eprint{1310.2536}.

\bibitem[{\citenamefont{Navarro~Perez
  et~al.}(2014{\natexlab{b}})\citenamefont{Navarro~Perez, Garrido, Amaro, and
  Arriola}}]{Perez:2014laa}
\bibinfo{author}{\bibfnamefont{R.}~\bibnamefont{Navarro~Perez}},
  \bibinfo{author}{\bibfnamefont{E.}~\bibnamefont{Garrido}},
  \bibinfo{author}{\bibfnamefont{J.}~\bibnamefont{Amaro}}, \bibnamefont{and}
  \bibinfo{author}{\bibfnamefont{E.~R.} \bibnamefont{Arriola}},
  \bibinfo{journal}{Phys.Rev.} \textbf{\bibinfo{volume}{C90}},
  \bibinfo{pages}{047001} (\bibinfo{year}{2014}{\natexlab{b}}),
  \eprint{1407.7784}.

\bibitem[{\citenamefont{Navarro~Perez
  et~al.}(2014{\natexlab{c}})\citenamefont{Navarro~Perez, Amaro, and
  Ruiz~Arriola}}]{Perez:2014jsa}
\bibinfo{author}{\bibfnamefont{R.}~\bibnamefont{Navarro~Perez}},
  \bibinfo{author}{\bibfnamefont{J.}~\bibnamefont{Amaro}}, \bibnamefont{and}
  \bibinfo{author}{\bibfnamefont{E.}~\bibnamefont{Ruiz~Arriola}},
  \bibinfo{journal}{Phys.Lett.} \textbf{\bibinfo{volume}{B738}},
  \bibinfo{pages}{155} (\bibinfo{year}{2014}{\natexlab{c}}),
  \eprint{1407.3937}.

\bibitem[{\citenamefont{Perez et~al.}(2015)\citenamefont{Perez, Nogga, Amaro,
  and Ruiz~Arriola}}]{Nogga-2015}
\bibinfo{author}{\bibfnamefont{R.~N.} \bibnamefont{Perez}},
  \bibinfo{author}{\bibfnamefont{E.}~\bibnamefont{Nogga}, \bibfnamefont{A.}},
  \bibinfo{author}{\bibfnamefont{J.}~\bibnamefont{Amaro}}, \bibnamefont{and}
  \bibinfo{author}{\bibfnamefont{E.}~\bibnamefont{Ruiz~Arriola}},
  \bibinfo{journal}{Work in preparation}  (\bibinfo{year}{2015}).

\bibitem[{\citenamefont{Caprio et~al.}(2012)\citenamefont{Caprio, Maris, and
  Vary}}]{Caprio:2012rv}
\bibinfo{author}{\bibfnamefont{M.}~\bibnamefont{Caprio}},
  \bibinfo{author}{\bibfnamefont{P.}~\bibnamefont{Maris}}, \bibnamefont{and}
  \bibinfo{author}{\bibfnamefont{J.}~\bibnamefont{Vary}},
  \bibinfo{journal}{Phys.Rev.} \textbf{\bibinfo{volume}{C86}},
  \bibinfo{pages}{034312} (\bibinfo{year}{2012}), \eprint{1208.4156}.

\bibitem[{\citenamefont{Sternberg et~al.}(2008)\citenamefont{Sternberg, Ng,
  Yang, Maris, Vary, Sosonkina, and Le}}]{sternberg2008accelerating}
\bibinfo{author}{\bibfnamefont{P.}~\bibnamefont{Sternberg}},
  \bibinfo{author}{\bibfnamefont{E.~G.} \bibnamefont{Ng}},
  \bibinfo{author}{\bibfnamefont{C.}~\bibnamefont{Yang}},
  \bibinfo{author}{\bibfnamefont{P.}~\bibnamefont{Maris}},
  \bibinfo{author}{\bibfnamefont{J.~P.} \bibnamefont{Vary}},
  \bibinfo{author}{\bibfnamefont{M.}~\bibnamefont{Sosonkina}},
  \bibnamefont{and} \bibinfo{author}{\bibfnamefont{H.~V.} \bibnamefont{Le}}, in
  \emph{\bibinfo{booktitle}{Proceedings of the 2008 ACM/IEEE conference on
  Supercomputing}} (\bibinfo{organization}{IEEE Press}, \bibinfo{year}{2008}),
  p.~\bibinfo{pages}{15}.

\bibitem[{\citenamefont{Maris et~al.}(2010)\citenamefont{Maris, Sosonkina,
  Vary, Ng, and Yang}}]{maris2010scaling}
\bibinfo{author}{\bibfnamefont{P.}~\bibnamefont{Maris}},
  \bibinfo{author}{\bibfnamefont{M.}~\bibnamefont{Sosonkina}},
  \bibinfo{author}{\bibfnamefont{J.~P.} \bibnamefont{Vary}},
  \bibinfo{author}{\bibfnamefont{E.}~\bibnamefont{Ng}}, \bibnamefont{and}
  \bibinfo{author}{\bibfnamefont{C.}~\bibnamefont{Yang}},
  \bibinfo{journal}{Procedia Computer Science} \textbf{\bibinfo{volume}{1}},
  \bibinfo{pages}{97} (\bibinfo{year}{2010}).

\bibitem[{\citenamefont{Aktulga et~al.}(2012)\citenamefont{Aktulga, Yang, Ng,
  Maris, and Vary}}]{aktulga2012topology}
\bibinfo{author}{\bibfnamefont{H.~M.} \bibnamefont{Aktulga}},
  \bibinfo{author}{\bibfnamefont{C.}~\bibnamefont{Yang}},
  \bibinfo{author}{\bibfnamefont{E.~G.} \bibnamefont{Ng}},
  \bibinfo{author}{\bibfnamefont{P.}~\bibnamefont{Maris}}, \bibnamefont{and}
  \bibinfo{author}{\bibfnamefont{J.~P.} \bibnamefont{Vary}}, in
  \emph{\bibinfo{booktitle}{Euro-Par 2012 Parallel Processing}}
  (\bibinfo{publisher}{Springer}, \bibinfo{year}{2012}), pp.
  \bibinfo{pages}{830--842}.

\bibitem[{\citenamefont{Navratil et~al.}(2009)\citenamefont{Navratil,
  Quaglioni, Stetcu, and Barrett}}]{Navratil:2009ut}
\bibinfo{author}{\bibfnamefont{P.}~\bibnamefont{Navratil}},
  \bibinfo{author}{\bibfnamefont{S.}~\bibnamefont{Quaglioni}},
  \bibinfo{author}{\bibfnamefont{I.}~\bibnamefont{Stetcu}}, \bibnamefont{and}
  \bibinfo{author}{\bibfnamefont{B.~R.} \bibnamefont{Barrett}},
  \bibinfo{journal}{J.Phys.} \textbf{\bibinfo{volume}{G36}},
  \bibinfo{pages}{083101} (\bibinfo{year}{2009}), \eprint{0904.0463}.

\bibitem[{\citenamefont{Gloeckner and Lawson}(1974)}]{gloeckner1974spurious}
\bibinfo{author}{\bibfnamefont{D.}~\bibnamefont{Gloeckner}} \bibnamefont{and}
  \bibinfo{author}{\bibfnamefont{R.}~\bibnamefont{Lawson}},
  \bibinfo{journal}{Phys.Lett.} \textbf{\bibinfo{volume}{B53}},
  \bibinfo{pages}{313} (\bibinfo{year}{1974}).

\bibitem[{\citenamefont{Forssén et~al.}(2008)\citenamefont{Forssén, Vary,
  Caurier, and Navratil}}]{Forssen:2008qp}
\bibinfo{author}{\bibfnamefont{C.}~\bibnamefont{Forssén}},
  \bibinfo{author}{\bibfnamefont{J.~P.} \bibnamefont{Vary}},
  \bibinfo{author}{\bibfnamefont{E.}~\bibnamefont{Caurier}}, \bibnamefont{and}
  \bibinfo{author}{\bibfnamefont{P.}~\bibnamefont{Navratil}},
  \bibinfo{journal}{Phys. Rev.} \textbf{\bibinfo{volume}{C77}},
  \bibinfo{pages}{024301} (\bibinfo{year}{2008}), \eprint{0802.1611}.

\bibitem[{\citenamefont{Maris and Vary}(2013)}]{Maris:2013poa}
\bibinfo{author}{\bibfnamefont{P.}~\bibnamefont{Maris}} \bibnamefont{and}
  \bibinfo{author}{\bibfnamefont{J.~P.} \bibnamefont{Vary}},
  \bibinfo{journal}{Int.J.Mod.Phys.} \textbf{\bibinfo{volume}{E22}},
  \bibinfo{pages}{1330016} (\bibinfo{year}{2013}).

\bibitem[{\citenamefont{Coon et~al.}(2012)\citenamefont{Coon, Avetian, Kruse,
  van Kolck, Maris et~al.}}]{Coon:2012ab}
\bibinfo{author}{\bibfnamefont{S.~A.} \bibnamefont{Coon}},
  \bibinfo{author}{\bibfnamefont{M.~I.} \bibnamefont{Avetian}},
  \bibinfo{author}{\bibfnamefont{M.~K.} \bibnamefont{Kruse}},
  \bibinfo{author}{\bibfnamefont{U.}~\bibnamefont{van Kolck}},
  \bibinfo{author}{\bibfnamefont{P.}~\bibnamefont{Maris}},
  \bibnamefont{et~al.}, \bibinfo{journal}{Phys.Rev.}
  \textbf{\bibinfo{volume}{C86}}, \bibinfo{pages}{054002}
  (\bibinfo{year}{2012}), \eprint{1205.3230}.

\bibitem[{\citenamefont{Furnstahl et~al.}(2012)\citenamefont{Furnstahl, Hagen,
  and Papenbrock}}]{Furnstahl:2012qg}
\bibinfo{author}{\bibfnamefont{R.}~\bibnamefont{Furnstahl}},
  \bibinfo{author}{\bibfnamefont{G.}~\bibnamefont{Hagen}}, \bibnamefont{and}
  \bibinfo{author}{\bibfnamefont{T.}~\bibnamefont{Papenbrock}},
  \bibinfo{journal}{Phys.Rev.} \textbf{\bibinfo{volume}{C86}},
  \bibinfo{pages}{031301} (\bibinfo{year}{2012}), \eprint{1207.6100}.

\bibitem[{\citenamefont{Wiringa et~al.}(1995)\citenamefont{Wiringa, Stoks, and
  Schiavilla}}]{Wiringa:1994wb}
\bibinfo{author}{\bibfnamefont{R.~B.} \bibnamefont{Wiringa}},
  \bibinfo{author}{\bibfnamefont{V.~G.~J.} \bibnamefont{Stoks}},
  \bibnamefont{and}
  \bibinfo{author}{\bibfnamefont{R.}~\bibnamefont{Schiavilla}},
  \bibinfo{journal}{Phys. Rev.} \textbf{\bibinfo{volume}{C51}},
  \bibinfo{pages}{38} (\bibinfo{year}{1995}), \eprint{nucl-th/9408016}.

\bibitem[{\citenamefont{Stoks et~al.}(1994)\citenamefont{Stoks, Klomp,
  Terheggen, and de~Swart}}]{Stoks:1994wp}
\bibinfo{author}{\bibfnamefont{V.~G.~J.} \bibnamefont{Stoks}},
  \bibinfo{author}{\bibfnamefont{R.~A.~M.} \bibnamefont{Klomp}},
  \bibinfo{author}{\bibfnamefont{C.~P.~F.} \bibnamefont{Terheggen}},
  \bibnamefont{and} \bibinfo{author}{\bibfnamefont{J.~J.}
  \bibnamefont{de~Swart}}, \bibinfo{journal}{Phys. Rev.}
  \textbf{\bibinfo{volume}{C49}}, \bibinfo{pages}{2950} (\bibinfo{year}{1994}),
  \eprint{nucl-th/9406039}.

\bibitem[{\citenamefont{Navarro~Pérez
  et~al.}(2013{\natexlab{c}})\citenamefont{Navarro~Pérez, Amaro, and
  Ruiz~Arriola}}]{Perez:2013mwa}
\bibinfo{author}{\bibfnamefont{R.}~\bibnamefont{Navarro~Pérez}},
  \bibinfo{author}{\bibfnamefont{J.~E.} \bibnamefont{Amaro}}, \bibnamefont{and}
  \bibinfo{author}{\bibfnamefont{E.}~\bibnamefont{Ruiz~Arriola}},
  \bibinfo{journal}{Phys. Rev.} \textbf{\bibinfo{volume}{C88}},
  \bibinfo{pages}{024002} (\bibinfo{year}{2013}{\natexlab{c}}),
  \bibinfo{note}{[Erratum: Phys. Rev.C88,no.6,069902(2013)]},
  \eprint{1304.0895}.

\bibitem[{\citenamefont{Navarro~Pérez
  et~al.}(2014)\citenamefont{Navarro~Pérez, Amaro, and
  Arriola}}]{Perez:2013oba}
\bibinfo{author}{\bibfnamefont{R.}~\bibnamefont{Navarro~Pérez}},
  \bibinfo{author}{\bibfnamefont{J.~E.} \bibnamefont{Amaro}}, \bibnamefont{and}
  \bibinfo{author}{\bibfnamefont{E.~R.} \bibnamefont{Arriola}},
  \bibinfo{journal}{Phys. Rev.} \textbf{\bibinfo{volume}{C89}},
  \bibinfo{pages}{024004} (\bibinfo{year}{2014}), \eprint{1310.6972}.

\bibitem[{\citenamefont{Audi et~al.}(2002)\citenamefont{Audi, Wapstra, and
  Thibault}}]{Audi:2002rp}
\bibinfo{author}{\bibfnamefont{G.}~\bibnamefont{Audi}},
  \bibinfo{author}{\bibfnamefont{A.~H.} \bibnamefont{Wapstra}},
  \bibnamefont{and} \bibinfo{author}{\bibfnamefont{C.}~\bibnamefont{Thibault}},
  \bibinfo{journal}{Nucl. Phys.} \textbf{\bibinfo{volume}{A729}},
  \bibinfo{pages}{337} (\bibinfo{year}{2002}).

\bibitem[{\citenamefont{Navarro~Perez
  et~al.}(2014{\natexlab{d}})\citenamefont{Navarro~Perez, Amaro, and
  Arriola}}]{Perez:2014waa}
\bibinfo{author}{\bibfnamefont{R.}~\bibnamefont{Navarro~Perez}},
  \bibinfo{author}{\bibfnamefont{J.~E.} \bibnamefont{Amaro}}, \bibnamefont{and}
  \bibinfo{author}{\bibfnamefont{E.~R.} \bibnamefont{Arriola}}
  (\bibinfo{year}{2014}{\natexlab{d}}), \eprint{1410.8097}.

\end{thebibliography}

\end{document}